\documentclass[letterpaper,american]{article}
\usepackage{jheppub}
\usepackage[T1]{fontenc}
\usepackage[latin9]{inputenc}
\usepackage{tipa}
\usepackage{tipx}
\usepackage{amsmath}
\usepackage{amssymb}

\makeatletter

\pdfpageheight\paperheight
\pdfpagewidth\paperwidth

\usepackage{bbm}

\makeatother

\usepackage{babel}
\begin{document}

\title{A holographic model for black hole complementarity}

\author[a]{David A. Lowe}
\emailAdd{lowe@brown.edu}

\affiliation[a]{Physics Department, Brown University, Providence, RI, 02912, USA}

\author[b,c]{Larus Thorlacius}
\emailAdd{lth@hi.is}
\affiliation[b]{University of Iceland, Science Institute, Dunhaga 3, IS-107, Rey
kjavik,
Iceland}
\affiliation[c]{The Oskar Klein Centre for Cosmoparticle Physics, Department of Physics, Stockholm
University, AlbaNova University Centre, 10691 Stockholm, Sweden}
\abstract{
We explore a version of black hole complementarity, where an approximate
semiclassical effective field theory for interior infalling degrees
of freedom emerges holographically from an exact evolution of exterior
degrees of freedom. The infalling degrees of freedom have a complementary
description in terms of outgoing Hawking radiation and must eventually
decohere with respect to the exterior Hamiltonian, leading to a breakdown
of the semiclassical description for an infaller. Trace distance is
used to quantify the difference between the complementary time evolutions,
and to define a decoherence time. We propose a dictionary where the
evolution with respect to the bulk effective Hamiltonian corresponds
to mean field evolution in the holographic theory. In a particular
model for the holographic theory, which exhibits fast scrambling,
the decoherence time coincides with the scrambling time. The results
support the hypothesis that decoherence of the infalling holographic
state and disruptive bulk effects near the curvature singularity are
complementary descriptions of the same physics, which is an important
step toward resolving the black hole information paradox. 
}
\maketitle

\section{Introduction}

The essence of the black hole information paradox is that the symmetry
principle of General Relativity, general covariance, is incompatible
with a unitary quantum evolution, where the Hawking radiation \citep{hawking}
carries away the quantum information of the black hole. The internal
rigidity of quantum mechanics, and the success of holographic approaches
to string theory, such as AdS/CFT \citep{Maldacena:1997re} and the
BFSS matrix model \citep{Banks:1996vh}, lend weight to the hypothesis
that a unitary quantum description should be exact. One is then faced
with the problem of how to recover approximate general covariance
from such a description.

According to the principle of black hole complementarity, as introduced
in \citep{Susskind:1993if}, physics outside the stretched horizon
of a black hole is well described by a local effective field theory
but the local description does not extend inside the stretched horizon.
As far as outside observers are concerned, the black hole interior
is encoded into quantum mechanical degrees of freedom associated with
the stretched horizon and residing in a Hilbert space of finite dimension
given by the exponential of the Bekenstein-Hawking entropy of the
black hole. No information enters the black hole - from the outside
point of view infalling matter is absorbed, thermalized and re-emitted
by the stretched horizon - and the interior spacetime experienced
by a typical observer entering the black hole in free fall is postulated
to emerge from the stretched horizon degrees of freedom in a holographic
fashion. The interior hologram is constructed from a finite number
of degrees of freedom, so the interior bulk theory can at best be
approximately local. The question is whether a physical observer inside
a black hole, whose measurement resolution is limited both in space
and time by the finite size of the black hole, can detect a deviation
from local effective field theory in the time allotted before hitting
the curvature singularity. 

A version of black hole complementarity, which addresses this question,
has been proposed by the authors and explored in some recent papers
\citep{Lowe:2014vfa,Lowe:2015eba}. The construction in \citep{Lowe:2014vfa}
applies to an observer falling into a black hole at a prescribed time
and is restricted to a limited time period before and after. For the
construction to work, the two complementary descriptions of the black
hole interior need to satisfy two key requirements. First of all,
from the point of view of the outside description, the minimal decoherence
time of a black hole must have a lower bound of order the scrambling
time $t_{scr}=4M\log(4M)$. This is the minimum time it takes outside
observers to extract quantum information from the black hole after
an infalling observer has been absorbed into the stretched horizon.
Second, from the viewpoint of the infalling observer, any quantum
information that entered the black hole more than a scrambling time
earlier must already have been erased from the interior bulk description
when the observer enters the black hole. 

The second requirement was studied in \citep{Lowe:2015eba} using
a simple model for the bulk physics. It was argued that a sensible
holographic description with a finite $N$ (or more precisely the
bulk $\hbar$ non-zero), would correspond to a bulk theory with a
physical regulator\footnote{The need for a bulk regulator to reconcile 
a finite black hole entropy with the number of states computed 
semiclassically was noted already in \citep{tHooft:1984re}.}. A minimal 
requirement is that such a theory should be capable of
describing events in a freely-falling frame outside the black hole
with a resolution down to a Planck length. Propagating that forward
in time, then leads to a lattice black hole model of the type studied
in \citep{Corley:1997ef} based on a Painlev\textipa{\'{e}}-Gullstrand
time-slicing. The time coordinate of the infalling frame can be matched
to the exterior timelike Killing vector at some finite distance outside
the black hole, allowing one to map time evolution in the holographic
model to time evolution in the black hole interior. By studying interior
propagation of massless fields in this lattice description, it was
found that the scrambling time emerges as the maximum coordinate time
a signal can propagate before hitting the curvature singularity. This
is in sharp contrast to the unregulated description, with exact general
covariance, where an infinite coordinate time might pass before collision
with the singularity.

In the present paper, we study a model for the holographic side of
this story. The predictions from the bulk side are that one should
see free propagation in the interior for a time at least of order
the scrambling time \citep{Lowe:2015eba} followed by the rapid onset
of large curvature effects with a timescale of order the Planck time.
Suppose we send in a small subsystem in a pure state into the black
hole. The subsystem can be viewed as a simple model for a freely falling
laboratory, where tests of spacetime locality are carried out. Eventually
the quantum information initially contained in the subsystem will
come out in the Hawking radiation. Since the Hawking radiation propagates
with respect to a local Hamiltonian in the exterior, any interactions
there will appear as effects that decohere the state from the interior
bulk viewpoint. Moreover, such interactions will appear highly non-local
in the infalling frame, and lead to apparent violations of quantum
mechanics for the infalling observer. Naively, one might predict that
measurements of the Hawking radiation might disrupt such a state
in an arbitrarily short time. However, if one insists that the detector
itself evolve according to the same Hamiltonian as the black hole
degrees of freedom, a finite minimal decoherence time emerges.

These statements will be quantified in a spin system we introduce
below as a simple model of the holographic stretched horizon. The
model we choose exhibits fast scrambling which is conjectured to be
a property of the stretched horizon degrees of freedom a wide class
of black holes \citep{Sekino:2008he}. The infalling Hamiltonian evolution
is mapped in the holographic model to a mean field Hamiltonian, dependent
on the initial state of the system. We compute the trace distance
between the states that are obtained by evolving the initial state
with respect to the exact Hamiltonian and with respect to the mean
field Hamiltonian. This trace distance provides a measure of the decoherence
of the infalling state. We find that the decoherence only becomes
significant after at least a scrambling time, matching precisely the
expectation from the regulated bulk theory. Moreover the timescale
for the rapid onset of decoherence also matches the bulk prediction.
The results support the version of black hole complementarity advocated
by the authors, where singularity approach is complementary to decoherence
of the infalling state, initially outlined in \citep{Lowe:2006xm}.
This represents an important step forward towards solving the black
hole information problem.

\section{Coherence/decoherence}

Let us begin by reviewing the basic ideas of decoherence, which involves
some system of interest $S$, interacting with some much larger system
$S^{C}$ which is often denoted the environment. We suppose the Hilbert
space factors as

\[
\mathcal{H}=\mathcal{H}_{S}\otimes\mathcal{H}_{S^{c}}
\]
Consider a pure state in $\mathcal{H}_{S}$
\[
|\Psi\rangle=\left(|\psi_{1}\rangle_{S}+|\psi_{2}\rangle_{S}\right)\otimes|\chi\rangle_{S^{C}}
\]
Under time evolution this becomes
\begin{equation}
|\Psi'\rangle=e^{-iHt}|\Psi\rangle=\sum_{i}c_{1i}|e_{i}\rangle\otimes|f_{1i}\rangle+c_{2i}|e_{i}\rangle\otimes|f_{2i}\rangle\label{eq:timeevo}
\end{equation}
where the $e_{i}$ are some basis of $\mathcal{H}_{S}$. If there
is decoherence, then it is a good approximation to assume $|f_{1i}\rangle$
is orthogonal to $|f_{2j}\rangle$ for \emph{any} $i$ and $j$. For
example, this will typically occur if the Hamiltonian is local in
position space and $|f_{1i}\rangle$ and $|f_{2i}\rangle$ are position
eigenstates. We will adopt the notation
\[
\Psi_{S}=\mathrm{Tr}_{S^{c}}|\Psi\rangle\langle\Psi|
\]
to denote the partial trace over the complement of $S$. If there
is decoherence, then to a good approximation
\begin{align}
\Psi'_{S} & \approx\sum_{i}\left(|c_{1i}|^{2}+|c_{2i}|^{2}\right)|e_{i}\rangle\langle e_{i}|\label{eq:diagden}
\end{align}
which means the probabilities add, without cross terms. The end result
for $\Psi'_{S}$ is then the same as if a measurement had collapsed
the wavefunction into the states $|e_{i}\rangle$. Note the probability
of each $|e_{i}\rangle$ is not necessarily equal, so $\Psi'_{S}$
need not be maximally mixed.

As it stands, this statement of decoherence is basis dependent. To
make a basis independent statement, one can instead quantify the purity
of the reduced density matrix $\Psi_{S}'$. One way to do this is to
compute
\[
P=\mathrm{Tr}_{S}(\Psi_{S}')^{2}
\]
which is known as the purity of a density matrix. $P=1$ for a pure
state since a pure state acts as a projector, $\left(\Psi_{S}'\right)^{2}=\Psi_{S}'$,
and by normalization $\mathrm{Tr}_{s}\Psi'_{S}=1$. For a mixed state
$0<P<1$. 

Alternatively, one may use the von Neumann entropy
\[
\mathcal{S}=-\mathrm{Tr}_{s}\Psi'_{S}\log\Psi_{S}'
\]
to quantify the purity of the reduced density matrix. This vanishes
for a pure state. For a maximally mixed state $\Psi_{S}'=\mathbbm{1}/n$,
on the other hand, $\mathcal{S}=\log n$, with $n$ the dimension of the 
Hilbert subspace $S$.

We can then formulate the decoherence time $t_{d}$ in the following
way. Assume at time $t=0$ $\Psi_{S}$ is in a pure state. Then define
the decoherence time $t_{d}$ as the time when
\begin{equation}
\mathcal{S}(\Psi_{S}(t_{d}))=\delta\log n\label{eq:decoentropy}
\end{equation}
for some choice of $\delta<1$. We are not aware of prior appearances
of this definition of decoherence time in the literature. This definition
should be useful in many other contexts.

In the following we will mostly be interested in studying finite dimensional
spin systems. In this class of models, we can reformulate the condition
\eqref{eq:decoentropy} as a condition on the trace distance using
the results of \citep{fannes1973}. We recall the definition 
\begin{equation}
\left\Vert \Psi_{S}-\Phi_{S}\right\Vert _{1}=\mathrm{Tr}_{S}\sqrt{\left(\Psi_{S}-\Phi_{S}\right)^{\dagger}\left(\Psi_{S}-\Phi_{S}\right)}\label{eq:tracedis}
\end{equation}
In \citep{fannes1973} it is shown that 
\begin{equation}
|\mathcal{S}(\Psi_{S})-\mathcal{S}(\Phi_{S})|\leq\left\Vert \Psi_{S}-\Phi_{S}\right\Vert _{1}\log n\label{eq:fannes}
\end{equation}
for two different density matrices in $\mathcal{H}_{S}$. Therefore
the definition of the decoherence time can be reformulated as
\begin{equation}
\left\Vert \Psi_{S}(t_{d})-\Phi_{S}(t_{d})\right\Vert _{1}=\delta\label{eq:tracedeco}
\end{equation}
for some fixed constant $\delta<1$ and some suitable choice for $\Phi_{S}$.
The state $\Phi_{S}(t)$ should be chosen to maintain purity under
time evolution for the subsystem of interest, but minimize the trace
distance as a function of time so the bound \eqref{eq:fannes} is
as useful as possible. For the models considered here, we will choose
$\Phi_{S}$ to evolve according to a local mean field Hamiltonian,
as we describe below.

In \citep{Lashkari:2011yi} the statement of fast scrambling was defined
in a similar way. The key distinction is that scrambling involves
a global mixing of the system, rather than only the mixing of a particular
subsystem of interest. The condition for scrambling would then require
that, \eqref{eq:tracedeco} should hold for all subsystems, suitably
defined, rather than some single small subsystem, as typically considered
in the decoherence problem.

\section{Toy Holographic Model\label{sec:Toy-Holographic-Model}}

While it is interesting to try to derive an effective holographic
model for the horizon degrees of freedom of a black hole from some
more fundamental description such as AdS/CFT or the Matrix Model,
our strategy will be to make some minimal assumptions about such a
description, and hope that it carries over to a more precise reconstruction.
The key assumption we will make of the model is that it exhibits fast
scrambling in the sense of \citep{Hayden:2007cs}, with a scrambling
time
\[
t\sim\beta\log S_{BH}
\]
with $\beta$ the inverse Hawking temperature of a black hole hole
with energy $E$ and $S_{BH}$ the Bekenstein-Hawking entropy of the
black hole. Later we will also be interested in carrying out computations
in the model for highly entangled states that will model the state
of an old black hole entangled with its Hawking radiation. As such,
we assume the model contains enough degrees of freedom to model the
interior of the black hole and its immediate vicinity. Thus we make
the identification that $S\sim N$ the number of sites in the model,
and $\beta$ will be scaled out of the problem. The near-horizon region
will not contain all the symmetries of the asymptotic region, so we
do not expect conformal symmetry (as in AdS/CFT) or supersymmetry
(as in the BFSS model) to be crucial in formulating this effective
model. At best the holographic model should contain a version of rotation/translation
symmetry, and time translation invariance.

A toy model that exhibits these features is discussed in \citep{Lashkari:2011yi}.
This is a spin model with a non-local pairwise interaction. There
are $N$ distinct sites with the Hilbert space of tensor product form
$\mathcal{H}=\mathcal{H}_{1}\otimes\cdots\otimes\mathcal{H}_{N}$.
The sites interact via a pairwise Hamiltonian 
$H=\sum_{\left\langle x,y\right\rangle }H_{\left\langle x,y\right\rangle }$
summing over unordered pairs of sites. The Hamiltonian may therefore
be associated with a graph $G=(V,E)$ with $N$ vertices $V$, and
edges $E$ corresponding to the non-zero $H_{\left\langle x,y\right\rangle }$.
In order to have fast scrambling, the degree of the vertices $D$
should be of order the size of the system. We shall then set $D=N-1$.
To have a sensible limit for large $N$, we take the pairwise interactions
to be bounded $\left\Vert H_{\left\langle x,y\right\rangle }\right\Vert<c/D$, for some
constant $c$. Here the operator norm $\left\Vert {\cal O} \right\Vert$ may be defined as the absolute value of the maximum eigenvalue of the operator ${\cal O}$.

The Lieb-Robinson result \citep{lieb1972} places bounds on the norm
of the commutator of operators localized at different sites, as a
function of time. For local interactions, this is to be interpreted
as a proof of finite group velocity in nonrelativistic spin systems.
In the case at hand, where interactions are non-local, the same method
still yields a bound on the norm of the commutator for operators.
In particular, in \citep{Lashkari:2011yi} it is shown that
\begin{equation}
\left\Vert \left[O_{A}(t),O_{B}\right]\right\Vert 
\leq\frac{4}{D}\left\Vert O_{A}\right\Vert \left\Vert O_{B}\right\Vert |A|e^{8ct}
\label{eq:liebrob}
\end{equation}
Here $O_{X}$ is a bounded norm operator acting in the Hilbert subspace
of the sites in the set $X$, and $B$ is chosen to be a single site.

The condition for scrambling is set up in \citep{Lashkari:2011yi} as
follows. Consider some Hilbert subspace $\mathcal{H}_{1}$ with dimension
of order 1, maximally entangled with some reference system $\mathcal{P}$,
which experiences no interactions. Here we set the system $S=\mathcal{H}_{1}$.
Under time evolution, the entanglement between $\mathcal{H}_{1}$
and $\mathcal{P}$ will decay, which may be quantified by the trace
distance
\begin{equation}
\left\Vert \Psi_{\mathcal{P}S}(t_{*})-\Psi_{\mathcal{P}}(0)
\otimes\Psi_{S}(t_{*})\right\Vert _{1}<\epsilon\mathrm{\,rank}\Psi_{\mathcal{P}}(0)
\label{eq:tracedist}
\end{equation}
for some constant $\epsilon\ll1$. This may in principle then be used
as a definition of scrambling time. A bound on the time $t_{*}$ 
can then be obtained by noting that it is bounded by the signaling
time from the space $S$ to its complement $S^{c}$, which may be
bounded using \eqref{eq:liebrob}.

First apply this to an initial state where $S$ is a single site,
and the complement subspace $S^{c}$ has dimension of order $N$. We assume
the initial state is of product form $|\Psi(0)\rangle=|\psi_{1}\rangle_{\mathcal{PH}_{1}}
\otimes|\psi_{2}\rangle_{\mathcal{H}_{2}}\otimes\cdots
\otimes|\psi_{N}\rangle_{\mathcal{H}_{N}}$.
Applying \eqref{eq:liebrob} with $B=S$ and $A=S^{c}$ one finds the timescale 
$t_{*}$ is of order a constant. 

For the black hole problem, the natural initial state to choose is
instead one where the black hole degrees of freedom are maximally
entangled with the exterior Hawking radiation. Now essentially the
roles of $S$ and $S^{c}$ are reversed. One takes the system $S$
to be of size of order $N$, with some small subsystem in a factor
pure state. The complement is then of size of order 1. To satisfy the
bound \eqref{eq:tracedist} one again requires signaling between $S$
and $S^{c}$, and this timescale is bounded by the Lieb-Robinson result.
This yields a timescale of order $\log N$ as expected for a fast
scrambling system.

\section{Mean field and bulk evolution\label{sec:Mean-field-and}}

At first sight, the results of the previous section are not encouraging
for the black hole complementarity scenario. While one can build holographic
models that exhibit fast scrambling with $t_{*}$ proportional to
$\log N$, it seems the decoherence time for some small Hilbert subspace
in such models will be very short. This is, however, not the right question
to ask in the black hole problem. Instead, what one should do is 
build a model for a laboratory that one sends into the black hole, and then
ask whether that laboratory will have a decoherence time sufficiently
long that they will not be able to distinguish quantum mechanics failing
from their classical demise due to singularity approach.

The eventual failure of quantum mechanics in the infalling laboratory 
can be traced to the existence of two distinct time evolutions for the
state in the lab subspace. One of these will be the exact Hamiltonian
evolution according to the holographic Hamiltonian $H$. The other 
will be defined according to a mean field Hamiltonian $H_{MF}$, 
that we describe in more detail shortly, and corresponds to the 
usual notion of time evolution in the bulk spacetime.
It is important to note that not all states yield sensible mean field
evolutions. Moreover, as will be clear, the mean field Hamiltonian
depends on the state. We conjecture that states close to smooth bulk
spacetimes do have useful mean field descriptions, and that the mean
field evolution is dual to the usual time evolution with respect to
the bulk Hamiltonian. In the remainder of this section, we explore
these issues in the context of our holographic toy model. 

The mean field approximation to the time evolution of a density matrix
is considered in some generality in \citep{muntean}. We begin by 
briefly reviewing the standard mean field construction based on an 
initial pure state of product form 
\begin{equation}
|\Psi(0)\rangle=|\psi_{1}\rangle_{\mathcal{H}_{1}}
\otimes\cdots\otimes|\psi_{N}\rangle_{\mathcal{H}_{N}}
\label{eq:initial}
\end{equation}
and later on we adapt it to the case of a highly entangled state 
corresponding to an old black hole. 
 
Starting from \eqref{eq:initial} one builds a state dependent mean field 
Hamiltonian 
\begin{align}
H^{MF} & =\sum_{x}H_{x}^{MF}(t)\nonumber \\
H_{x}^{MF} & =\sum_{y}{\rm tr}_{y}
\left(H_{\left\langle x,y\right\rangle }\Psi_{y}^{MF}(t)\right)
\label{eq:meanham}
\end{align}
where $\Psi^{MF}$ evolves according to $H^{MF}$ from the
same initial state $|\Psi(0)\rangle$. A key point is that with these
definitions, and choice of initial state, the mean field Hamiltonian
never generates entanglement between different sites, and remains
in the same product form as the initial state. This mean field Hamiltonian
then has the expected properties of the holographic dual of the bulk
gravity Hamiltonian. As is well known, the bulk Hamiltonian generates
smooth time evolution all the way to the curvature singularity, with
minimal entanglement being generated. This feature is in fact the origin 
of the information problem.\footnote{It should be noted the mean field 
Hamiltonian depends on the choice
of initial state via \eqref{eq:meanham}. The state dependence of
the boundary to bulk map is emphasized in \citep{Papadodimas:2013jku}.
However in the present construction it is also important that the boundary
to bulk mapping is time dependent, which follows from $[H,H_{x}^{MF}]\neq0.$}

One then wishes to calculate the timescale for which the trace norm
distance between $\Psi_{x}(t)$ and $\Psi_{x}^{MF}(t)$ remains small.
This maps onto a problem solved in \citep{Lashkari:2011yi} for the spin model
considered above, via careful application of Lieb-Robinson bounds
applied to an expansion of the matrix element
\begin{equation}
\left\langle \Psi(t)|\mathbbm{1}-\Psi_{x}^{MF}(t)|\Psi(t)\right\rangle 
=1-\left\langle \Psi_{x}^{MF}(t)|\Psi_{x}(t)|\Psi_{x}^{MF}(t)\right\rangle 
\label{eq:matrixel}
\end{equation}
by making a Dyson series expansion in $H-H_{x}^{MF}$. This matrix
element in turn places a bound on the trace distance between the states
\eqref{eq:tracedis}. Using the result of \citep{fannes1973}, this
then places a bound on the von Neumann entropy $H(\Psi_{x}(t))$.
One finds
\begin{equation}
\left\langle \Psi_{x}^{MF}(t)|\Psi_{x}(t)|\Psi_{x}^{MF}(t)
\right\rangle \leq\frac{c'}{D}e^{c''t}
\label{eq:liebrob-lak}
\end{equation}
where $c'$and $c''$ are constants independent of $N$. For $D=N-1$
these quantities become of order 1 when $t\sim\log N$.

Making contact with black hole physics, the holographic description
should be useful both inside and outside the black hole horizon. An
initial state of the form \eqref{eq:initial} is relevant outside
the black hole horizon, or for a recently formed black hole prior
to scrambling. To make further progress we need to generalize the
mean field results of \citep{Lashkari:2011yi} to highly entangled
states.

Suppose we choose a maximally entangled initial state where we have
a pairwise entanglement between $\mathcal{H}_{2k}$ and $\mathcal{H}_{2k+1}$
for all $k\geq1$. Then we can almost map the problem into the one
just considered by coarse graining, and viewing 
$\mathcal{H}_{2k}\otimes\mathcal{H}_{2k+1}$
as a pure state on a single coarse grained site.\footnote{We note that 
the Schmidt decomposition \citep{Nielsen:2011:QCQ:1972505} implies this 
special state is unitary equivalent to the generic maximally entangled state. 
Converting from Schrodinger picture to Heisenberg picture, this may be 
viewed as conjugation of the Hamiltonian by a constant unitary matrix. 
For the pairwise interaction considered in this model, a general unitary 
transformation will induce self-interactions, but preserve the pairwise form 
of the Hamiltonian.  The condition $\left\Vert H_{\langle x,y\rangle}\right\Vert<c/D$ 
is preserved by this unitary conjugation, so the above proof goes through.} 
The new feature is that the coarse grained Hamiltonian now has a
self-interaction term. Such a term must be treated exactly in the
mean field approximation. For this initial state, we therefore define
\begin{align*}
H^{MF} & =\sum_{x}H_{x}^{MF}(t)\\
H_{x}^{MF} & =H_{\left\langle x,x\right\rangle }
+\sum_{y\backslash x}{\rm tr}_{y}
\left(H_{\left\langle x,y\right\rangle }\Psi_{y}^{MF}(t)\right)
\end{align*}
where the sums are over coarse grained sites $x=1,\cdots,N/2$. With
this Hamiltonian, we may then proceed as above to compute the trace
distance between the mean field state and the exact evolution, or
equivalently the von Neumann entropy of the exact reduced density
matrix, obtaining the same scaling with $N$ (though different constants
$c'$ and $c''$) via \eqref{eq:liebrob-lak}. 

This is now a nice model for an old evaporating black hole after the
Page time $t\sim\beta S_{BH}$, where the interior degrees of freedom
are maximally entangled with the exterior Hawking radiation. The decoherence
time, defined according to the definition \eqref{eq:decoentropy}
is now of order $t\sim\log N$ matching the scrambling time. We also
note once a time of order the scrambling time has passed, the bound
\eqref{eq:liebrob-lak} increases with a rise time of order 1, matching
the bulk expectation of strong curvature effects with an onset of
order the Planck time.

The same kind of computations can be carried out for a variety of
initial states. For example, to mimic a an observer falling in and
carrying out quantum experiments, we can choose to separate the infalling
site $x$ into two sites $x_{1}$ and $x_{2}$ with 
$\left\Vert \mathcal{H}_{1}\Vert\gg \Vert\mathcal{H}_{2}\right\Vert$.
Let us assume some strong coupling between these sites, with coupling
to the other sites bounded as above. Defining the decoherence time
using \eqref{eq:tracedeco} for the choice $S=x_{2}$ will lead to
a decoherence time of order 1, independent of $N$. This means the
decoherence time for measurements internal to the infalling lab can
be made rapidly, as expected. However it is still true that the combined
state on $\mathcal{H}_{1}\otimes\mathcal{H}_{2}$ will remain pure
for a time of order the scrambling time using the above construction.
Thus measurements of the Hawking radiation will not lead to rapid
decoherence of the state, as naively expected.

To test the idea that the minimal decoherence time due to measurement
of Hawking radiation really matches the scrambling time, one can try
to generalize the above discussion to any density matrix with a pure
state subfactor representing the infalling system. The above derivation
will generalize provided the mean field approximation holds for the
evolution of the subsystem of interest. It seems natural that states
representing smooth spacetime geometries will correspond to good mean
field states, however the converse need not be true. It would be very
interesting to see more directly how this class of states emerges
as a class of attractor states from large $N$ holographic theories.

\section{Comments}

In the above we have argued the minimal decoherence time of an infalling
state is to be identified with the scrambling time, subject to assumptions
about the form of the holographic model. In this context, we have
proven one of the key assumptions about the approach to black hole
complementarity described in \citep{Lowe:2014vfa,Lowe:2015eba}. 

The other key assumption relied on details of the holographic reconstruction
of bulk spacetimes, namely that general covariance is softly broken
through the introduction of a Planck length regulator. This was explored
in a regulated model of the bulk in \citep{Lowe:2015eba}, where it
was found that it was sufficient for infalling degrees of freedom
to retain coherence for a scrambling time. That work describes the
details of the construction, including the important conditions placed
on bulk timeslices compatible with the regulator.

Since we are concluding the timescales match, this is an important
success for building interior degrees of freedom in a holographic
theory. Moreover, if the bound is saturated, one also predicts the
timescale of order 1 associated with the rapid rise expected from
strong curvature near the singularity. Away from this region the trace
distance will be of order $1/N$.

Since these corrections are to be essentially interpreted as violations
of quantum mechanics for the infalling observer, it is of great interest
to quantify to what extent these are tolerable. The trace distance
may be interpreted directly as the probability of an ideal experiment
detecting the difference between two states \citep{Nielsen:2011:QCQ:1972505}.
As an initial crude estimate, if we take $N\sim S_{universe}\approx10^{88}$
(assuming domination by CMB photons) and assume the nonlocal effect
produces a Planck energy particle with a probability $1/N$ per unit
Planck time per degree of freedom, we can apply it to the atoms in
the Earth's atmosphere to conclude one Planck energy ultra high energy
cosmic ray would appear about every $10^{7}$ years. This is conceivably
a detectable effect, but apparently rather harmless. 

It is natural to conjecture some version of the same matching of scrambling
time with interior decoherence time holds in all holographic theories
of quantum gravity. It remains an important open problem to directly
derive holographic effective theories of the horizon degrees of freedom
from more fundamental descriptions such as AdS/CFT or Matrix Models,
and test this conjecture. It will also be very interesting to further
explore mean field approximations in such holographic descriptions.
Of course one expects the mean field (or master field formulation
of a large $N$ theory) will coincide with the bulk gravity description
at leading order. However having a formulation directly in the Hilbert
space of the underlying holographic description is needed to carry out
computations analogous to those of section~\ref{sec:Mean-field-and}.

We note that generic holographic states in more realistic models may
well contain singularities leading to a breakdown of the mean field approach.  From the bulk
perspective, we would expect for sufficiently large $N$, a version of cosmic 
censorship \citep{Penrose:1969pc} to hold. The additional singularities 
will then be censored by their own horizons leaving a smooth geometry 
outside where we expect mean field to remain accurate. It will be interesting to see what extent this may be derived in the holographic model considered here.

In the near term, it would be interesting to generalize the present
work beyond spin models to systems with an infinite dimensional Hilbert
space at each site. Some recent applications of Lieb-Robinson bounds
to such systems appear in \citep{Nachtergaele2009,Nachtergaele:2009jv,2014arXiv1410.8174N,2015arXiv151206319G}.
It appears promising that these results can be generalized to models
that exhibit fast scrambling.

\appendix

\section{Comparison with firewall story}

It is useful to compare the black hole complementarity picture with
the popular firewall story.

\subsection{Disagreement in overlap region}

The formulation of \citep{Lowe:2014vfa,Lowe:2015eba} implies that the infalling state begins to disagree with
the exact state outside the black hole a scrambling time
prior to horizon crossing. One might attempt to reach a contradiction
with this approach by taking an outgoing Hawking particle emitted
in this overlap region, performing a quantum computation on this,
and the earlier Hawking radiation, and sending the result to the infalling
observer. If this could be done with appreciable probability, it would
result in a violation of the ordinary rules of quantum mechanics for
the infaller, who would notice that upon horizon crossing, the required
entangled Hawking partner was not there.

If the unitary transformation is done using the holographic model
we have described above, the quantum computation maps to unitary evolution
with respect to some dense pairwise interactions in the holographic
theory. The above estimates apply to this state as well, and one again
concludes that the time necessary for such a computation is of order
the scrambling time. Thus by construction, the infaller has already
crossed the horizon.

\subsection{Precomputation}

One might try an extreme version of the above by precomputing the
quantum state, and arranging for the infalling state to be entangled
with the outgoing Hawking particle in the overlap region. In this
case the black hole complementarity picture fails, and is unpredictive
for the experience of the infalling observer.

However we should then try to quantify how surprising it is for the
approach to fail for special states. To arrange for one Hawking particle
to be precisely entangled with a given infaller requires picking a
vector lying in an $e^{S_{BH}}$ dimensional Hilbert space, modulo
unitary transformations that just act on the other Hawking particles.
Standard estimates, matching the trace distance of such states to
the required state then give the time for such a transformation to
be constructed of order
\[
e^{ce^{S_{BH}(M)}}/e^{ce^{S_{BH}(M-1/\beta)}}\approx e^{c'e^{S_{BH}(M)}}
\]
where $c$ and $c'$ are positive constants of order 1. This is parametrically
the same as upper bounds on the quantum Poincare recurrence time \citep{Page:1994dx,2015arXiv150904352C}.

We conclude that with enough available time, one can reliably create
infalling states that do not have sensible bulk evolution in the interior.
However unless one completely isolates the state, for an extremely
long time, any quantum noise will defeat the coherence required to
see these unusual effects. Turning the argument around, a typical
infalling observer will see drama with a probability of order $e^{-c'e^{S_{BH}}}$.

\subsection{Microcanonical ensemble}

In \citep{Marolf:2013dba} it was suggested the microcanonical ensemble
would lead to an infalling number operator that always was of order
1 for any mode, due to entanglement with the exterior operators. Loopholes
in this argument were already pointed out in \citep{Avery:2013vwa,Avery:2015hia}
and the present work and its companion paper \citep{Lowe:2015eba}
provide a concrete realization of these ideas. The new ingredients
are the time-dependent (with respect to CFT time) bulk regulator \citep{Lowe:2015eba}
and the corresponding time and state dependent boundary to bulk map
via the mean field Hamiltonian. 

In particular, if one were to replace the mean field evolution by
evolution with respect to a state independent operator a decoherence
time of order 1 will emerge for typical states according to \eqref{eq:tracedeco}.
For example if we tried to represent the bulk Hamiltonian by the identity
operator, this would correspond to the choice $\Phi_{S}(t)=\Psi_{S}(0)$
in \eqref{eq:tracedeco}. For the kinds of highly entangled states
considered in section~\ref{sec:Mean-field-and} this will lead to
a timescale of order 1. The mean field approach is essential for correctly
matching the holographic dual of the bulk Hamiltonian.
\begin{acknowledgments}
D.L. wishes to thank Daniel Harlow for a helpful discussion. The research
of D.L. was supported in part by DOE grant de-sc0010010. The research
of L.T. was supported in part by Icelandic Research Fund grant 163422-051,
the University of Iceland Research Fund, and the Swedish Research
Council under contract 621-2014-5838.
\end{acknowledgments}

\bibliographystyle{jhep}
\bibliography{trace}

\end{document}